\documentclass[final]{hubsarxiv}

\usepackage{siunitx}
\usepackage{longtable}
\usepackage{flafter}
\usepackage{placeins}
\usepackage{etoolbox}

  \pretocmd{\section}{\FloatBarrier}{}{}
  \pretocmd{\subsection}{\FloatBarrier}{}{}

\usepackage{tabularx}
\newcolumntype{Y}{>{\raggedright\arraybackslash}X}

  \newcommand{\traceref}{Appendix~\ref{app:trace}}

\graphicspath{{figures/}}

\newcommand{\papertitle}{The one-step surface--air lag in the EnergyPlus zone heat balance: order of
  accuracy, correction and consequences}
\newcommand{\paperabstract}{The order of accuracy sets how fast numerical error falls as the timestep shortens. This study
measured that order for the zone heat balance of EnergyPlus 26.2 against exact or converged
references and tested a correction on 45 U.S. Department of Energy prototype cases. All three zone
air formulas, the third-order default included, give first order or lower, because the surfaces use
the previous timestep's air temperature. Solving surfaces, air and HVAC to agreement within each
timestep raises the default to about second order, and in a room with an exact solution two extra
passes cut the error eightfold at 1.65 times the runtime. In the prototype cases, the correction
changed annual site energy by at most 0.45\% but an outpatient clinic's heating by 5.1\%.
Window-upgrade savings changed by 10\% or more in 13 of the 34 applicable cases, about as often as
under a change of timestep.}
\newcommand{\paperkeywords}{EnergyPlus\kwsep heat balance method\kwsep numerical accuracy\kwsep
  time integration\kwsep verification\kwsep building energy simulation}
\newcommand{\txtdepartment}{Department of Civil and Architectural Engineering and Mechanics}
\newcommand{\txtcompeting}{The author reports there are no competing interests to declare.}
\newcommand{\txtfunding}{No funding was obtained for this work.}
\newcommand{\txtavailability}{The modification of EnergyPlus, as patch files against release
  26.2.0, the scripts that generate and run the input files of the tests on simple models and the
  prototype study, the outcome measures and thresholds set before the prototype runs, and the
  summary results are openly available at
  \url{https://github.com/humanbuildingsynergy/EnergyPlus-SurfaceAirCoupling} and archived at
  \url{https://doi.org/10.5281/zenodo.24041718}. EnergyPlus 26.2.0 and the ASHRAE 90.1-2019
  prototype models are publicly available from their developers.}

\newcommand{\txtaiuse}{During the preparation of this work the author used Claude Opus 5.5
  (Anthropic) to support writing and checking the simulation and analysis scripts, including the
  modification of the EnergyPlus source code, and to draft and edit the text. After using this tool,
  the author reviewed and edited the content as needed and takes full responsibility for the content
  of the publication.}
\newcommand{\txtcredit}{\textbf{Wooyoung Jung:} Conceptualization, Methodology, Software, Formal
  analysis, Investigation, Visualization, Writing -- original draft, Writing -- review \&
  editing.}

  \newcommand{\kwsep}{;}
  \title{\papertitle}
  \shorttitle{The surface--air lag in the EnergyPlus heat balance}
  \shortauthors{Jung}
  \author{Wooyoung Jung\affmark{a,}\corrmark\orcid{0000-0003-2746-4983}}
  \affil{\affmark{a}\txtdepartment, University of Arizona, Tucson, AZ, USA}
  \corresponding{wooyoung@arizona.edu}
  \paperno{2026-010}
  \paperdate{November 2026}
  \versionnote{v1 (2026-11-07): first release.}
  \license{CC BY-NC-ND 4.0}
  \abstract{\paperabstract}

\begin{document}

  \maketitle

\section{Introduction}\label{sec:intro}
Building energy simulation is used to predict how much energy a building uses and, as buildings are
increasingly asked to shift or shed load in response to the grid~\cite{neukomm2019geb}, when it uses
it. Such predictions must resolve the building's response within the day. Simulation programs
advance the heat balances of a building in discrete timesteps, and results within the day are more
sensitive to the length of the timestep than annual totals are~\cite{tabaresvelasco2013timestep}. A
shorter timestep reduces the numerical error at a cost in runtime, and how much it reduces the error
depends on the order of accuracy of the solution, the rate at which the error falls as the timestep
is shortened. When the timestep is halved, the error of a first-order solution halves, whereas that
of a third-order solution falls eightfold, so the order determines how much runtime a given accuracy
within the day requires.

EnergyPlus~\cite{crawley2001energyplus}, the program behind the prototype buildings of the U.S.
Department of Energy and the national building stock
models~\cite{goel2014prototypes,parker2023comstock,reyna2025resstock}, makes a third-order formula
the default for updating the zone air temperature, alongside the first-order Euler method and an
analytical solution described as free of truncation error~\cite{eplus2026engref,taylor1991impact}.
Within each timestep, however, the program calculates the surfaces from the air temperature of the
previous timestep~\cite{eplus2026source}, and the theory of coupled problems predicts that such a
one-step lag limits the whole solution to first order whatever formula updates the
air~\cite{arnold2013error,trcka2010cosim}. A shorter timestep would then buy accuracy only at the
first-order rate.

Whether this prediction holds for EnergyPlus has not been tested. Studies of EnergyPlus have found
that its results depend strongly on the timestep. In an early version of the program, moving a large
office from a 15-minute to an hourly timestep raised heating, ventilation and air-conditioning
(HVAC) electricity use by 50.9\% and heating gas use by 28.5\%~\cite{hong2008runtime}, and in
high-performance houses with pre-cooling, hourly and one-minute timesteps gave different afternoon
cooling loads~\cite{tabaresvelasco2013timestep}. These studies report how the outputs of particular
buildings change, without separating the numerical error of the solution from other effects of the
timestep, such as the resolution of controls. The verification of simulation programs leaves the
question open as well. Comparative tests check whether programs agree with one
another~\cite{judkoff1995bestest,ashrae2023std140,ashrae2025std140adda}, and analytical tests
compare parts of a model, such as heat conduction through the building fabric, with exact
solutions~\cite{spitler2001analytical,tabaresvelasco2012diagnostic}. Neither kind measures the order
of accuracy of the whole solution, in which surfaces, zone air and systems exchange heat.

This study measures that order for the zone heat balance of EnergyPlus, finds where the error
arises, and examines whether correcting it changes the results of models of real buildings, and at
what cost in runtime. It asks three research questions (RQs).
\begin{itemize}
  \item \textbf{RQ1, accuracy.} How fast does the error of the zone heat balance fall as the
    timestep is shortened, and does any of the three zone air schemes reach the order of accuracy
    that its formula promises?
  \item \textbf{RQ2, cause and correction.} Where in the solution does this error arise, and does
    solving the surface, air and system balances to agreement within each timestep remove it? At
    what cost in runtime?
  \item \textbf{RQ3, consequences.} In the prototype buildings of the U.S. Department of Energy,
    does removing the error change annual energy, peak demand, unmet hours, equipment sizes or the
    savings of a measure, and does it reach a given accuracy in the hourly zone air temperature at
    less runtime than a shorter timestep?
\end{itemize}
Three hypotheses, drawn from the order of calculation in the program and from the theory of coupled
problems, guide the analysis.
\begin{itemize}
  \item \textbf{H1 (RQ1).} Because the surface heat balances use the zone air temperature of the
    previous timestep, the solution would be only first order with all three schemes, the
    third-order and analytical schemes included.
  \item \textbf{H2 (RQ2).} Because this lag would be the source of the first-order error, solving
    the balances to agreement within each timestep would raise the order of the default scheme
    above one, toward the third order of its formula.
  \item \textbf{H3 (RQ3).} Because the lag acts in every timestep of every zone, removing it would
    materially change annual heating, cooling or fan energy, peak electric demand, unmet hours,
    equipment sizes or the savings of a window upgrade in at least a quarter of the cases, and the
    corrected program would reach the accuracy of the release at less than half its runtime.
\end{itemize}

The study modifies the source code of EnergyPlus so that the surface, air and system balances can
be brought into agreement within each timestep, and compares the modified program with the
released one. For RQ1 and RQ2, simple models separate the numerical solution from the complexity of
real buildings: a single room run with each scheme at 1 to 60 timesteps per hour, a wall compared
with its exact conduction solution, and a room reduced until its equations can be solved outside
the program. For RQ3, the 16 prototype buildings are run in three climates with both programs,
with outcomes and thresholds set before these simulations were run.

The study contributes a measured order of accuracy for the zone heat balance of EnergyPlus, which
tells users what a shorter timestep buys, and a correction that developers can adopt as an option.
The changes to the source code and the scripts of every test are released, so the measurements can
be repeated for later versions of the program and the same tests applied to other programs.

\section{Background and related work}\label{sec:background}
\subsection{The zone air heat balance}\label{sec:bg-zone}

EnergyPlus represents the air of each zone as one well-mixed volume at temperature $T_z$. Its energy
balance is~\cite{eplus2026engref}
\begin{equation}
  \begin{aligned}
    C_z \frac{dT_z}{dt} = {} & \sum_i \dot{Q}_i + \sum_i h_i A_i \left(T_{si} - T_z\right)
      + \sum_i \dot{m}_i c_p \left(T_{zi} - T_z\right) \\
    & + \dot{m}_{inf} c_p \left(T_\infty - T_z\right) + \dot{Q}_{sys},
  \end{aligned}
  \label{eq:zone}
\end{equation}
where $C_z$ is the heat capacity of the zone air, $\dot{Q}_i$ are convective internal gains, $h_i
A_i (T_{si} - T_z)$ is convection from surface $i$ at temperature $T_{si}$, the next two terms are
air exchange with other zones and with the outdoors, and $\dot{Q}_{sys}$ is the heat supplied by the
HVAC system. The surface temperatures come from a heat balance on each face of each surface, which
includes conduction through the construction, radiation and convection to the zone
air~\cite{eplus2026engref}. The air and surface balances are therefore coupled through $T_z$ and
$T_{si}$.

The HVAC system enters through a predictor--corrector procedure~\cite{eplus2026engref}. The program
first computes the heating or cooling that would hold the zone at its setpoint. It then simulates
the system to find what it can deliver, and finally updates $T_z$ with the delivered amount. Within
a zone timestep, the HVAC calculation may use shorter system timesteps~\cite{eplus2026ioref}.

Three schemes advance Eq.~\eqref{eq:zone} in time~\cite{eplus2026engref}. The user selects one in
the \texttt{ZoneAirHeatBalanceAlgorithm} object, with the keys \texttt{EulerMethod},
\texttt{ThirdOrderBackwardDifference} (the default) and
\texttt{AnalyticalSolution}~\cite{eplus2026ioref}. The Euler method approximates the derivative by
the change in $T_z$ over one step divided by the step length $\delta t$, and it is first order in
time~\cite{eplus2026engref,taylor1991impact}. The default scheme uses the third-order backward
difference
\begin{equation}
  \frac{dT_z}{dt} \approx \frac{1}{\delta t}\left(\frac{11}{6}T_z^{t} - 3T_z^{t-\delta t}
  + \frac{3}{2}T_z^{t-2\delta t} - \frac{1}{3}T_z^{t-3\delta t}\right),
  \label{eq:bdf3}
\end{equation}
where the superscripts denote the current and the three previous timesteps. The analytical scheme
integrates Eq.~\eqref{eq:zone} exactly over the step with the other terms held constant, and it is
described as free of truncation error~\cite{eplus2026engref}.

\subsection{The lagged coupling and its origin}\label{sec:bg-lag}

\begin{figure}[tbp]
  \centering
  \includegraphics[width=\textwidth]{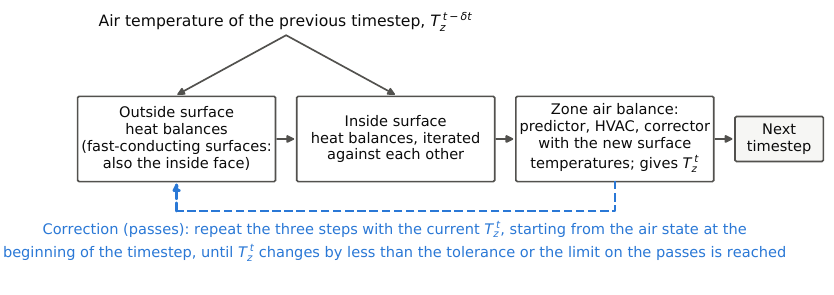}
  \caption{Order of calculation within a zone timestep in EnergyPlus, and the passes of the
  correction (dashed).}
  \label{fig:order}
\end{figure}

Within each zone timestep (Figure~\ref{fig:order}), EnergyPlus solves the heat balances on the
outside faces of the surfaces, then those on the inside faces, and then the zone air balance with
its predictor, HVAC simulation and corrector~\cite{eplus2026source}. The inside balances take the
zone air temperature from the end of the previous timestep, as the air temperature in the convection
term and for the interior convection coefficients, and hold it fixed while they iterate the surfaces
against one another. For surfaces that conduct heat quickly, such as layers without thermal mass,
the outside balance includes the inside face and reads the same temperature. The zone air balance
then uses the surface temperatures just calculated, and the zone air temperature is updated only at
its end. The surfaces are calculated again only in zones with a radiant system, baseboard heater,
cooling panel, swimming pool or surface-coupled photovoltaic panel, and then only once. The source
locations of these steps in release 26.2.0 are listed in \traceref{}. Every earlier release from
8.2.0 to 26.1.0 calculates in the same order.

This order goes back to BLAST (Building Loads Analysis and System Thermodynamics), one of the two
programs on which EnergyPlus was built~\cite{crawley2001energyplus}. Its solution loop iterated the
outside and inside surface balances against one another and solved the zone air balance after that
loop, and EnergyPlus kept a solution method and iteration scheme ``substantially the same as in
BLAST''~\cite{strand2001modularization}. The coupling of the zone air with the HVAC system comes
from IBLAST, a research version of BLAST~\cite{eplus2026engref}, in which Taylor et al.\ let the
system response lag the zone conditions by one step, which ``completely eliminates iteration from
the solution procedure'', and chose the third-order form among forms of second to fifth
order~\cite{taylor1991impact}.

The Engineering Reference describes the lagged coupling in deriving the update formula, where the
terms on the right side of Eq.~\eqref{eq:zone} ``are not known at the current time'' and ``were
lagged by one time step''~\cite{eplus2026engref}. It also repeats the argument of Taylor et al.,
credited to a personal communication, that the daily cycle of building simulation should make
truncation errors cancel over each cycle~\cite{taylor1991impact,eplus2026engref}. The input guide
notes that shorter timesteps improve how the surface and air temperatures are coupled, and that
longer timesteps ``introduce more lag''~\cite{eplus2026ioref}.

\subsection{Why a lag limits the order of accuracy}\label{sec:bg-order}

The order of a formula holds when every other quantity it uses is exact. In a model whose parts are
solved one after another, a part that uses values the other produced one step earlier sees inputs
that are off by about the change over one step, and the error this adds shrinks only in proportion
to the timestep. For coupled problems split into parts that exchange values held constant over each
step, this coupling error is of first order in the step length, whatever methods the parts
use~\cite{arnold2013error}, and in the co-simulation of a building and its HVAC system it reduces a
second-order method to first order~\cite{trcka2010cosim}.

The same reasoning applies inside EnergyPlus. Consider a surface without thermal mass. Its heat
balance then gives its temperature directly as a function of the air temperature, $T_{si} =
S(T_z)$, and Eq.~\eqref{eq:zone} can be written $C_z\,dT_z/dt = F\bigl(T_z, S(T_z)\bigr)$, where
$F$ stands for its right side. EnergyPlus evaluates the surface with the air temperature of the
previous step, so the right side becomes $F\bigl(T_z^{t}, S(T_z^{t-\delta t})\bigr)$. Because
$T_z^{t-\delta t}$ differs from $T_z^{t}$ by about $\delta t\,dT_z/dt$, this equals, to first order
in $\delta t$,
\begin{equation}
  F\bigl(T_z, S(T_z)\bigr)
  - \frac{\partial F}{\partial S}\,\frac{dS}{dT_z}\,\frac{dT_z}{dt}\,\delta t.
  \label{eq:lagerror}
\end{equation}
The first term is the true right side. The second is an extra heat flow, proportional to the
timestep and to how fast the air temperature changes. Because it is proportional to $dT_z/dt$, it
acts on the air like an extra heat capacity, $\sum_i h_i A_i\,(dS_i/dT_z)\,\delta t$, which grows
with the timestep. The program therefore solves a slightly different equation, and however
accurately it solves that equation, its result differs from the true solution by an amount
proportional to $\delta t$, whichever of the three schemes is used. For a surface with mass the same
kind of error arises, but it is smaller, because such a surface follows the air less closely.

\subsection{Coupling in other programs}\label{sec:bg-other}

Other building simulation programs resolve the coupling within the timestep. TRNSYS iterates its
connected components by successive substitution~\cite{wetter2006modelica}. ESP-r solves the energy
balances of the building simultaneously with a semi-implicit scheme that is second order in
time~\cite{clarke2001energy,clarke2004cooperating}, and IDA ICE and the Modelica Buildings library
solve the equations together, with a general-purpose solver~\cite{sahlin2004ida} or with implicit
algorithms of variable step size~\cite{wetter2006modelica,wetter2014modelica}. DOE-2 calculates the
load of each space at a constant temperature and corrects it in its systems
calculation~\cite{doe22basics}. Spawn of EnergyPlus moves the room air and the HVAC system to
Modelica, in part so that the fast dynamics of room air can use variable time step
solvers~\cite{wetter2020spawn}.

Where separately solved models are coupled, iterating within the timestep has proved more accurate
than using the other model's results from the previous step. For the heat and airflow models of
ESP-r, the lagged strategy with a six-minute timestep was less accurate in an atrium than iteration
with an hourly one, and it used more than twice the computer resources~\cite{hensen1999coupled}. In
co-simulation, strong coupling ``allows the use of longer time steps for the same accuracy'' than
loose coupling~\cite{trcka2009cosim}, and the loss of order under loose coupling has been measured
for research codes~\cite{gasparin2018improved}. Without iteration, extrapolating the exchanged
values from the previous steps can restore the order. Polynomial extrapolation of degree $k$ gives
order $k+1$ when the coupled system has no algebraic loops~\cite{arnold2013error}. In EnergyPlus the
lag lies inside the program, between the surface and air balances of every zone, and no option
removes it~\cite{eplus2026source}.

\section{Methods}\label{sec:methods}
\subsection{A coupled solution of the surface and air balances}\label{sec:m-coupled}

The source code of EnergyPlus 26.2.0, the version distributed by its developers and called the
release here, was modified so that the surface and air balances can be brought into agreement within
each zone timestep.\footnote{The patch files, the scripts and the results of this study are  available at
\url{https://github.com/humanbuildingsynergy/EnergyPlus-SurfaceAirCoupling} and archived at
\url{https://doi.org/10.5281/zenodo.24041718}.} After the usual sequence of a timestep, the
modified program solves the outside and inside surface heat balances again with the new zone air
temperature. It then returns the zone air and thermostat temperatures to their values at the start
of the timestep and repeats the load prediction, the HVAC simulation and the update of the zone air
temperature. One such repetition is called a pass (Figure~\ref{fig:order}, dashed). The passes stop
after a set number, or earlier when the zone air temperature changes by less than \SI{e-9}{\celsius}
between passes. The passes run in every zone timestep, including the design days used to size
equipment. The conduction histories, the history terms of the third-order formula and the
temperatures of water heater and plant loop tanks still advance once per timestep, and meters and
reports are accumulated once, after the passes.

Within its own calculation, the inside surface balance iterates the surfaces against one another
until their temperatures change by less than \SI{0.002}{\celsius}, a tolerance that the input file
cannot lower~\cite{eplus2026source}. Four configurations were used (Table~\ref{tab:configs}). The
two-pass configuration is the correction examined for practical use, and the converged configuration
serves as a reference. With the passes switched off, the modified program reproduced the outputs of
the release exactly, for a single room under all three zone air schemes and over a year for the
small office example file distributed with EnergyPlus, and it left the equipment sizing reports
unchanged~\cite{eplusaudit2026data}. The release was the Linux version distributed by the
developers, and the modified program was compiled from the same source code with the GNU C++
compiler (GCC) 13.3.

\begin{table}[htbp]
  \centering
  \caption{Program configurations compared in this study.}
  \label{tab:configs}
  \begin{tabularx}{\textwidth}{@{}lYl@{}}
    \toprule
    Configuration & Extra passes per timestep & Inside surface tolerance \\
    \midrule
    Release & none (EnergyPlus 26.2.0 as distributed) & \SI{0.002}{\celsius} \\
    One pass & 1 & \SI{0.002}{\celsius} \\
    Two passes & up to 2 & \SI{0.002}{\celsius} \\
    Converged & up to 50, stopping at \SI{e-9}{\celsius} & \SI{e-8}{\celsius} \\
    \bottomrule
  \end{tabularx}
  \par\smallskip\raggedright
  Pass: the outside and inside surface balances, the load prediction, the HVAC simulation and the
  zone air temperature calculated again within the timestep. Inside surface tolerance: the change in
  surface temperature at which the iteration of the inside surfaces stops.
\end{table}

\subsection{Tests on simple models}\label{sec:m-known}

These tests address hypotheses H1 and H2. Exploratory runs of the single room preceded them. The
order of accuracy was measured as the observed order of verification
studies~\cite{roache1998verification}. At each timestep $\delta t$, the error of a temperature is
the root mean square (RMS) difference between the computed values and the reference over the $N$
points at which both are available,
\begin{equation}
  e(\delta t) = \sqrt{\frac{1}{N}\sum_{k=1}^{N}\bigl(T_k(\delta t) - T_k^{\mathrm{ref}}\bigr)^2}.
  \label{eq:rms}
\end{equation}
The measured order $p$ is the slope of the straight line fitted by least squares to $\log e(\delta
t)$ against $\log \delta t$. Between two neighbouring timesteps, the local order $p_{\mathrm{loc}}$
is the slope between their two points. A reference run at a shorter timestep carries an error of its
own, which biases these slopes upward, so the order was also estimated without a reference, as
$p_3$, from runs at timesteps $\delta t$, $\delta t/2$ and $\delta
t/4$~\cite{roache1998verification}:
\begin{equation}
  p_{\mathrm{loc}} = \frac{\ln\bigl(e(\delta t_1)/e(\delta t_2)\bigr)}{\ln(\delta t_1/\delta t_2)},
  \qquad
  p_3 = \log_2 \frac{d(\delta t, \delta t/2)}{d(\delta t/2, \delta t/4)},
  \label{eq:orders}
\end{equation}
where $d$ is the RMS difference between two runs at the points they share. For the daily heating or
cooling energy of the ideal air system, the error is the difference from the reference divided by
the reference.

\begin{figure}[h]
  \centering
  \includegraphics[width=\textwidth]{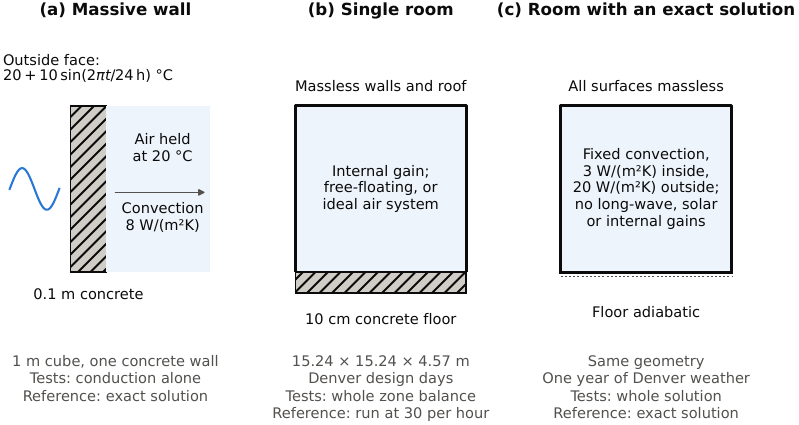}
  \caption{The three simple models, with what each tests and the reference it is compared with.}
  \label{fig:models}
\end{figure}

An error needs a reference. A run with a much shorter timestep is the usual choice, but in
EnergyPlus it is not a clean reference for constructions with mass. All surfaces in the tests and in
the prototype buildings use conduction transfer functions (CTFs), the program's default method for
heat conduction~\cite{eplus2026ioref}, which it derives from a finite-difference model of each layer
with node spacing $\sqrt{2\alpha\,\delta t}$, $\alpha$ being the thermal diffusivity, and 6 to 19
nodes per layer~\cite{eplus2026source}. A shorter timestep can therefore also change the spatial
grid. For a \SI{10}{cm} concrete layer, the grid has 6 nodes at every timestep up to 30 per hour and
9 nodes at 60 per hour. Each of the three tests below therefore uses either an exact solution or a
reference run on the same grid (Figure~\ref{fig:models}).

\paragraph{Conduction through a massive wall} A \SI{1}{m} cubic zone has one wall of a single
\SI{0.1}{m} layer (conductivity \SI{1.7}{W/(m.K)}, density \SI{2240}{kg/m^3}, specific heat
\SI{840}{J/(kg.K)}), and its other five surfaces are massless and adiabatic. The temperature of the
outside face is imposed as $20 + 10\sin(2\pi t/\SI{24}{h})$\,\si{\celsius}, and the inside face
exchanges heat with a fixed convection coefficient of \SI{8}{W/(m^2.K)} with zone air held at
\SI{20}{\celsius} by an ideal air system. Absorptances of $10^{-4}$ make long-wave and solar
exchange negligible. The steady periodic temperature of this slab is known in closed
form~\cite{carslaw1959conduction}. The inside face temperature of the release was compared with it
at the 24 hour ends, after at least 20 days of warmup, at 1 to 60 timesteps per hour.

\paragraph{A single room on design days} The single-zone example file distributed with EnergyPlus
(\texttt{1ZoneUncontrolled}) describes a zone of $15.24 \times 15.24 \times \SI{4.572}{m}$ with
massless walls and roof, a \SI{10}{cm} concrete floor, no windows and a constant internal gain. It
was run on its two Denver design days, each repeated until the solution was periodic, floating
without HVAC, with the solar gain removed, or with the floor replaced by a massless layer of equal
thermal resistance. With an ideal air system, the zone was held at \SI{22}{\celsius}, between 21 and
\SI{24}{\celsius} from 08:00 to 18:00 and between 15 and \SI{30}{\celsius} otherwise, or in a band
of 21 to \SI{22}{\celsius}, or stepped from 22 to \SI{20}{\celsius} from 08:00 to 18:00 with and
without the subdivision of the HVAC timestep. Each case ran with each of the three zone air schemes
at 1, 2, 3, 4, 5, 6, 10, 12, 15, 20, 30 and 60 timesteps per hour, with the release and the
converged configuration. Zone air and inside surface temperatures, and the daily heating or cooling
energy of the controlled cases, were compared at the 24 hour ends, which all timesteps share, with
the run at 30 timesteps per hour of the same configuration and scheme, which shares the CTF grid.
Orders were fitted from 2 to 10 timesteps per hour, and estimated without a reference from the runs
at 3, 6 and 12 and at 5, 10 and 20.

Further runs checked that the lag, and no other term, sets the order. They fixed the interior
convection coefficient at \SI{3}{W/(m^2.K)}, switched off interior long-wave exchange, suppressed
the subdivision of the HVAC timestep, or extended the warmup to 100 days. Runs that calculated only
the inside surfaces again separated the two places in which the lag occurs.

\paragraph{A room with an exact solution} The same example file was reduced until its equations
could be solved outside EnergyPlus to any desired accuracy. The floor was made massless, the
convection coefficients were fixed at \SI{3.0}{W/(m^2.K)} inside and \SI{20.0}{W/(m^2.K)} outside,
long-wave and solar absorptances were set near zero, the internal gains were removed, and the
outdoor temperature was made uniform with height. Each surface temperature then follows
algebraically from the air and outdoor temperatures, and the zone air obeys
\begin{equation}
  V \rho c_p \frac{dT}{dt} = UA \left(T_o(t) - T\right), \qquad
  UA = \sum_i \frac{A_i}{1/h_o + R_i + 1/h_i},
  \label{eq:exactroom}
\end{equation}
with $V = \SI{1061.88}{m^3}$ and $UA = \SI{144.0}{W/K}$, which gives a time constant of about two
hours. The density $\rho$ and specific heat $c_p$ of the air are evaluated with the psychrometric
functions of EnergyPlus, and the outdoor temperature $T_o$ and pressure are the hourly values of the
Denver weather file interpolated linearly between hour ends, as EnergyPlus interpolates
them~\cite{eplus2026source}. The humidity ratio, which EnergyPlus lowers when the zone air reaches
saturation, is taken from the converged configuration with 60 timesteps per hour and changes $\rho
c_p$ by about 0.06\%.

Equation~\eqref{eq:exactroom} was integrated over the year with an explicit Runge--Kutta method of
order eight~\cite{hairer1993solving} at relative and absolute tolerances of $10^{-12}$, and checked
with an implicit Radau method~\cite{hairer1996solving}, which agreed within \SI{2.1e-8}{\celsius}.
As in the warmup of EnergyPlus, the first day was repeated until periodic, and the first two days
were excluded from the comparison. The zone air temperature at the end of each zone timestep was
compared with the exact solution for all four configurations at the twelve timesteps listed above,
and orders were fitted from 4 to 20 timesteps per hour.

The limits on the passes were compared by their difference from the converged solution in the
free-floating single room on its summer design day at four timesteps per hour, and by their runtime
in an annual run of the small office example file (Table~\ref{tab:passes}).

\subsection{The prototype study}\label{sec:m-proto}

The consequences for practice were examined on the commercial prototype buildings of the U.S.
Department of Energy~\cite{deru2011prototypes,goel2014prototypes}, in their versions for ASHRAE
Standard 90.1-2019~\cite{pnnl2023prototypes}. All 16 buildings were taken in three climates, Miami
(climate zone 1A), Denver (5B) and International Falls (7), with the weather files distributed with
the models. The files, written for EnergyPlus 22.1, were converted to 26.2 with the transition
programs distributed with the release, and the only other change was a request for database output.
All files use six timesteps per hour and size their equipment automatically. The small office was
excluded, as set in advance, because its input file from the prototype set failed in all three
climates with both programs. Each of the remaining 45 cases was run with all four configurations,
and the change in its results was measured from the release to two passes.

Table~\ref{tab:thresholds} lists the outcome measures and the change that counts as material in a
case. These measures and thresholds were set before any prototype simulation was
run~\cite{eplusaudit2026data}, and they form the first part of hypothesis H3. A measure was judged
to matter in practice when at least a quarter of the eligible cases changed materially. The verdicts
rest on these thresholds and on no statistical test, because each simulation is deterministic, so a
difference between two runs contains no sampling noise. Unmet hours are small whole numbers, so they
have both a relative and an absolute threshold. The window upgrade sets every simple glazing system
whose U-factor and solar heat gain coefficient are both above \SI{1.42}{W/(m^2.K)} and 0.25 to those
values. Its savings are the annual site energy of the base case minus that of the upgraded case, and
cases in which no glazing qualified were excluded from this measure.

After the results were known, the verdicts were checked in further ways. A Wilcoxon signed-rank
test~\cite{wilcoxon1945individual}, used descriptively, asked whether each outcome changed in a
consistent direction across its eligible cases, and the counts of material changes were repeated
with every threshold halved and doubled. For the window upgrade, the change in savings was compared
with the spread of the savings among the three corrected configurations, which differ only in how
far the coupling is converged, and with the change when the release is run at four or ten timesteps
per hour. At ten per hour the outpatient clinic in International Falls stopped with a sizing error,
so that comparison has 33 cases. For the three cases with the largest changes in annual energy,
hourly outputs for every heating coil and fan located the change.

\begin{table}[htbp]
  \centering
  \caption{Outcome measures and the change that counts as material in a case.}
  \label{tab:thresholds}
  \begin{tabularx}{\textwidth}{@{}YY@{}}
    \toprule
    Outcome & Material change \\
    \midrule
    Annual heating, cooling or fan energy & at least 1\% of that end use \\
    Annual peak electric demand & at least 2\% \\
    Occupied unmet hours & at least 10\% and \SI{5}{h} for heating or cooling, or the total
      crossing \SI{300}{h} \\
    Autosized capacity of cooling coils, heating coils and supply fans, summed per building &
      at least 2\% \\
    Savings of a window upgrade & at least 10\%, or a change of sign \\
    \bottomrule
  \end{tabularx}
  \par\smallskip\raggedright
  A case changes materially when the change from the release to two passes reaches the threshold.
  Heating, cooling and fan energy count only where the end use is at least 5\% of the building's
  site energy. 300~h: the limit on unmet hours in the performance rating method of Standard
  90.1~\cite{ashrae2019std901}.
\end{table}

The second part of H3 asks whether the coupled solution reaches a given accuracy at less runtime
than the release. It was tested on three buildings in Denver, the medium office, the primary school
and the standalone retail building, each run at 4, 6, 10, 15, 20, 30 and 60 timesteps per hour with
the release, one pass and two passes. The reference was the converged configuration with 60
timesteps per hour, checked against its run with 30. The main error measure was the RMS difference
in hourly mean zone air temperature over all conditioned zones and all hours of the year, the
quantity on which the lag acts directly, and the errors in annual heating and cooling energy were
secondary. The coupled solution was to be judged faster if, in every building, the release needed at
least twice the runtime of two passes at four timesteps per hour to reach the same error, with the
runtime of the release interpolated between timesteps on logarithmic scales. All runs of the study
used one server (AMD EPYC 9554, 64 cores, 256~GB of memory, Ubuntu 24.04 under the Windows Subsystem
for Linux). The timed runs were executed one at a time, three times each, and the median runtime is
reported. Because each run sizes its equipment at its own timestep, the test was repeated after the
results were known with the sizes fixed at those of the release at six timesteps per hour, for the
release, two passes and the converged configuration at 4 to 30 timesteps per hour.

\section{Results}\label{sec:results}
\subsection{Order of accuracy of the release}\label{sec:r-h1}

\begin{figure}[tbp]
  \centering
  \includegraphics[width=\textwidth]{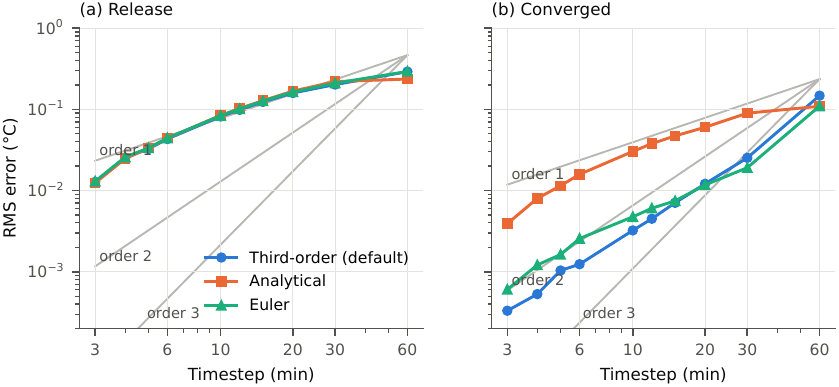}
  \caption{Error of the zone air temperature at the hour ends against the run with 30 timesteps per
  hour, for the single room on its summer design day, with (a) the release and (b) the surfaces and
  air calculated again until they agree. Gray lines show the slopes of orders 1, 2 and 3.}
  \label{fig:room}
\end{figure}

In the single room on its summer design day, the error of the zone air temperature falls at most in
proportion to the timestep with all three schemes (Figure~\ref{fig:room}a). The measured order is
0.96 for the default third-order formula, 0.97 for the Euler method and 1.00 for the analytical
solution. Estimated from three runs alone, each with half the timestep of the one before, the order
of the three schemes is 0.66 to 0.71 from the runs at 3, 6 and 12 timesteps per hour and 0.85 to
0.86 from those at 5, 10 and 20. It approaches one as the timestep shortens and stays below it. At
four timesteps per hour the errors of the three schemes are 0.124, 0.127 and \SI{0.128}{\celsius},
so the default formula is no more accurate than the Euler method, and the analytical solution is no
better. The inside surface temperatures follow the same pattern, with orders of 0.97 to 0.99.

Conduction alone does not explain this. For the massive wall, the conduction solution converges at
order 2.04 against a run on the same grid, and against the exact solution its error levels off at
\SI{0.023}{\celsius} from 10 to 30 timesteps per hour, the spatial error of the six-node grid.

\begin{figure}[tbp]
  \centering
  \includegraphics[width=0.6\textwidth]{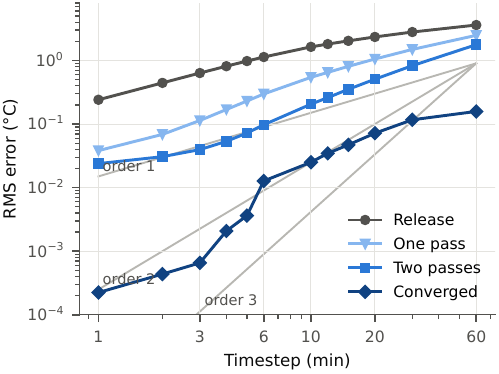}
  \caption{Error of the zone air temperature against the exact solution for the room without mass,
  with the default third-order formula, for the release and for the surfaces and air calculated
  again once, twice, or until they agree. Gray lines show the slopes of orders 1, 2 and 3.}
  \label{fig:exactroom}
\end{figure}

The room with an exact solution confirms the result against a reference computed outside EnergyPlus
(Figure~\ref{fig:exactroom}). The error of the release falls at order 0.73 from 4 to 20 timesteps
per hour and 0.56 from 2 to 10. Its local order rises from 0.37 between one and two timesteps per
hour to 0.88 between 30 and 60, toward first order. At six timesteps per hour, the rate of the
prototype models, the error of the release is \SI{1.63}{\celsius} RMS and \SI{3.9}{\celsius} at
most. This room has no thermal mass, so its surfaces follow the air directly and the lag is at its
largest.

\subsection{Solving the balances to agreement}\label{sec:r-h2}

With the surfaces and air in agreement, halving the timestep cuts the error of the default formula
about fourfold, against twofold with the release (Figure~\ref{fig:room}b). It follows the line of
order 2, below the order 3 of the formula alone. In the single room on its summer design day, the
measured order rises from 0.96 to 1.88, or to 1.97 and 1.63 when estimated from three runs alone,
and the error at four timesteps per hour falls from 0.124 to \SI{0.0071}{\celsius}, about seventeen
times smaller. With only the inside faces calculated again, the order stays at 0.97, because the
outside faces of the massless walls and roof still read the previous air temperature
(Table~\ref{tab:variants}). Once both faces are calculated again, no other term holds the solution
at first order. Fixing the interior convection coefficients, switching off interior long-wave
exchange or extending the warmup leaves the order at 1.4 or above in every estimate.

\begin{table}[tbp]
  \centering
  \caption{Order of accuracy and error of the zone air temperature in the single room without solar gain floating on its summer design day, with the default third-order scheme, for variants that change one term of the model.}
  \label{tab:variants}
  \begin{tabularx}{\textwidth}{@{}>{\raggedright\arraybackslash\leftskip=1em}p{0.37\textwidth}*{4}{>{\centering\arraybackslash}X}@{}}
    \toprule
    \multicolumn{1}{@{}l}{Variant} & Fitted order & \multicolumn{2}{c}{Order without a reference} & RMS error (\si{\celsius}) \\
    \cmidrule(lr){3-4}
    \multicolumn{1}{@{}l}{\emph{Timesteps per hour}} & 2--10, against 30 & 3, 6, 12 & 5, 10, 20 & 4 \\
    \midrule
    \multicolumn{5}{@{}l}{\emph{Release}} \\
    As described & 0.95 & 0.57 & 0.90 & 0.0492 \\
    Interior convection fixed at \SI{3}{W/(m^2.K)} & 0.95 & 0.57 & 0.84 & 0.0548 \\
    Interior long-wave exchange off & 1.24 & 1.06 & 0.97 & 0.0601 \\
    Both of the above & 0.98 & 0.66 & 0.71 & 0.0373 \\
    100 warmup days & 0.94 & 0.59 & 0.85 & 0.0487 \\
    \addlinespace
    \multicolumn{5}{@{}l}{\emph{Inside faces calculated again until the surfaces and air agree}} \\
    As described & 0.97 & 0.61 & 0.84 & 0.0420 \\
    Interior convection fixed at \SI{3}{W/(m^2.K)} & 0.96 & 0.60 & 0.83 & 0.0433 \\
    Interior long-wave exchange off & 1.00 & 1.70 & 0.82 & 0.0204 \\
    Both of the above & 0.97 & 0.61 & 0.83 & 0.0305 \\
    100 warmup days & 0.97 & 0.62 & 0.83 & 0.0420 \\
    \addlinespace
    \multicolumn{5}{@{}l}{\emph{Both faces calculated again until the surfaces and air agree}} \\
    As described & 1.85 & 1.85 & 1.73 & 0.0025 \\
    Interior convection fixed at \SI{3}{W/(m^2.K)} & 1.66 & 1.77 & 1.41 & 0.0022 \\
    Interior long-wave exchange off & 2.86 & 5.72 & 2.21 & 0.0016 \\
    Both of the above & 2.18 & 2.37 & 1.99 & 0.0015 \\
    100 warmup days & 1.85 & 1.85 & 1.73 & 0.0025 \\
    \bottomrule
  \end{tabularx}
  \par\smallskip\raggedright
  Order 1: the error halves when the timestep is halved. Order 2: it falls fourfold.
\end{table}

The other two schemes gain less. The Euler method reaches order 1.25, with an error of
\SI{0.0075}{\celsius} at four timesteps per hour. The analytical solution stays near first order, at
1.06, although its error falls from 0.128 to \SI{0.047}{\celsius}. In the room with an exact
solution, the Euler method gains as in the single room, from 2.06 to \SI{0.22}{\celsius} at four
timesteps per hour, but the analytical solution does not gain at all, with an error of 1.46 against
\SI{1.56}{\celsius} for the release. A calculation of its update at agreement, done outside
EnergyPlus, reproduces this error to \SI{0.0002}{\celsius}, so it belongs to the scheme and not to
the modified program.

In the room with an exact solution (Figure~\ref{fig:exactroom}), the error of the converged
configuration at six timesteps per hour is \SI{0.025}{\celsius}, 65 times smaller than that of the
release. Two passes bring it to \SI{0.21}{\celsius}, eight times smaller, at order 1.4, because they
leave part of the lag at coarse timesteps, and one pass to \SI{0.54}{\celsius}. The error of the
converged configuration follows no single order, 1.4 from 2 to 10 timesteps per hour and 2.6 from 4
to 20, because part of the lag remains. When the air temperature changes by more than
\SI{0.3}{\celsius} in a timestep, the program subdivides the HVAC timestep and advances the air in
the subdivisions with the surfaces held fixed~\cite{eplus2026source}. Without the subdivision, the
converged update calculated outside EnergyPlus has an error of \SI{0.0051}{\celsius} at four
timesteps per hour, against \SI{0.047}{\celsius} in the program.

In controlled zones, calculating the surfaces and air again with the HVAC held at its first result
moves the air off its setpoint by up to \SI{1.8e-3}{\celsius}, because the system delivers the load
predicted with the lagged surfaces. With the HVAC in the loop, the zone sits on its setpoint
whenever it is controlled. At four timesteps per hour, the daily cooling energy then differs from
the run with 30 timesteps per hour by 0.28\% with a thermostat that is always active, against 4.3\%
with the release, and by 0.61 to 0.77\% against 14.9 to 17.4\% when the zone floats within a band or
switches between floating and control. Steps of the setpoint are the weak point. A step of 2~K
triggers the subdivision of the HVAC timestep, and the error then no longer falls steadily as the
timestep shortens. With the subdivision suppressed, the measured order is 1.69.

Most of the gain comes with few passes (Table~\ref{tab:passes}). One extra pass per timestep leaves
a difference of \SI{0.023}{\celsius} from the converged solution at four timesteps per hour, against
\SI{0.148}{\celsius} for the release, at 1.35 times the runtime of the release. Two passes leave
\SI{0.005}{\celsius}, less than the remaining timestep error of the converged solution, at 1.65
times the runtime.

\begin{table}[tbp]
  \centering
  \caption{Mean extra passes per timestep, difference from the converged solution and runtime for each limit on the passes.}
  \label{tab:passes}
  \begin{tabularx}{\textwidth}{@{}Ycccc@{}}
    \toprule
    Limit & Extra passes & Difference (\si{\celsius}) & Runtime (s) & Ratio to release \\
    \midrule
    One pass & 1 & 0.0227 & 18.3 & 1.35 \\
    Two passes & 2 & 0.0050 & 22.4 & 1.65 \\
    Passes until the air changes by less than \SI{0.001}{\celsius} & 1.9 & 0.0059 & 26.3 & 1.93 \\
    Converged & 6.7 & -- & 79.5 & 5.85 \\
    \bottomrule
  \end{tabularx}
  \par\smallskip\raggedright
  Passes and difference: single room, free-floating, summer design day, four timesteps per hour. For comparison, the converged solution differs from its own run with 30 timesteps per hour by \SI{0.0071}{\celsius}, and the release differs from that run by \SI{0.148}{\celsius}. Runtime: annual run of the small office distributed with EnergyPlus, release 13.6~s, one sample each. Extra passes: the mean number of additional calculations of the surfaces, air and HVAC per timestep.
\end{table}

The daily cycle does not cancel the error of the release. At 30 timesteps per hour, its daily
cooling energy with a thermostat that is always active is still 1.1\% below the solution with the
balances in agreement, 7.16 against 7.24~kWh.

\subsection{Consequences in the prototype buildings}\label{sec:r-h3}

In the 45 prototype cases, two passes change the annual site energy by at most 0.45\%. Heating,
cooling or fan energy changes materially in 8 cases, 18\%, below the quarter that would make the
change matter in practice (Table~\ref{tab:h3}). Peak electric demand and unmet hours change
materially in one case each, and the autosized capacities in none. Across the eligible cases, the
median change in the three end uses is 0.19 to 0.31\%, and the largest is 5.1\%, in the heating of
the outpatient clinic in Miami. Peak demand changes by at most 2.1\% and the autosized capacities by
at most 1.9\%. The material change in unmet hours is in the large hotel in Miami, whose occupied
cooling unmet hours fall from 78.5 to 69.7~h. The converged configuration reached its limit of 50
passes in 3\% of the timesteps of the median case and in 39 to 75\% of those of the hospital and the
outpatient clinic, where on/off equipment switches state between passes. Across all cases, the
annual results of the converged configuration agree with those of two passes within 0.14 percentage
points, except for peak demand.

\begin{table}[tbp]
  \centering
  \caption{Cases in which each outcome changed materially, with each corrected configuration, and the verdict on hypothesis H3 for two passes.}
  \label{tab:h3}
  \begin{tabularx}{\textwidth}{@{}Yccccc@{}}
    \toprule
    Outcome & Cases & One pass & Two passes & Converged & Supported \\
    \midrule
    Annual heating, cooling or fan energy & 45 & 8 & 8 & 7 & no \\
    Annual peak electric demand & 45 & 1 & 1 & 1 & no \\
    Occupied unmet hours & 45 & 2 & 1 & 1 & no \\
    Autosized capacity & 45 & 0 & 0 & 0 & no \\
    Savings of the window upgrade & 34 & 11 & 13 & 12 & yes \\
    Accuracy at half the runtime & 3 buildings & -- & 0 & -- & no \\
    \bottomrule
  \end{tabularx}
  \par\smallskip\raggedright
  Cases: the eligible cases. One pass, two passes, converged: the cases that change materially, with the thresholds of Table~\ref{tab:thresholds}. Supported: at least a quarter of the cases change materially with two passes. Accuracy at half the runtime: the buildings in which the release needs at least twice the runtime of two passes at four timesteps per hour to reach their error in hourly air temperature.
\end{table}

\begin{figure}[tbp]
  \centering
  \includegraphics[width=\textwidth]{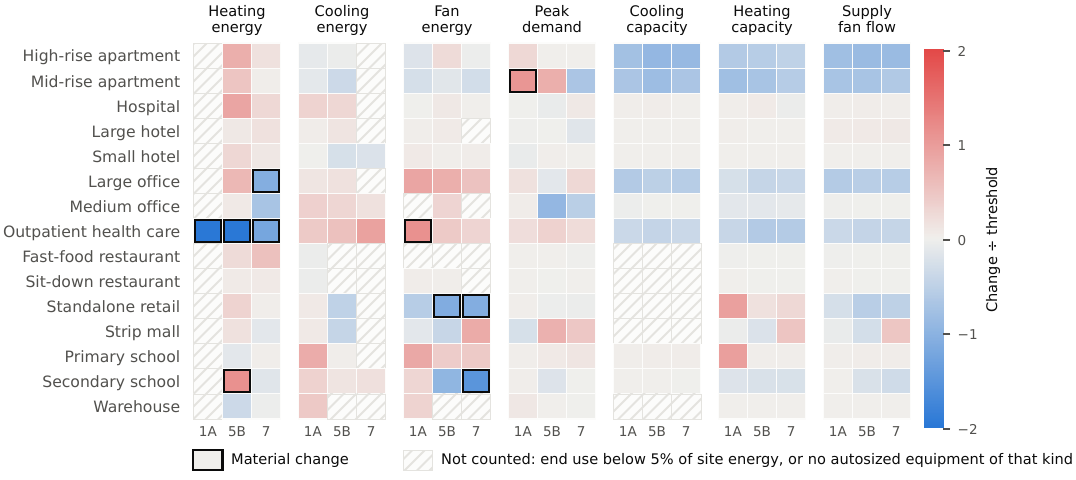}
  \caption{Change from the release to two passes in every prototype case, relative to the threshold
  for a material change, by outcome and climate zone (1A Miami, 5B Denver, 7 International Falls).
  Red marks an increase and blue a decrease, and the colors saturate at twice the threshold.}
  \label{fig:case-heatmap}
\end{figure}

The material changes cluster in a few buildings (Figure~\ref{fig:case-heatmap}). The outpatient
clinic changes in heating in all three climates, by $-5.1$, $-2.9$ and $-1.2$\%, and in Miami also
in fan energy. Its heating falls in the reheat coils of the variable air volume boxes by a similar
amount in every month, including the Miami summer, when the reheat coils keep the minimum air flow
of the operating rooms at temperature. The secondary school changes in heating in Denver and in fan
energy in International Falls, where three of its fans carry most of the drop.

\begin{figure}[tbp]
  \centering
  \includegraphics[width=0.6\textwidth]{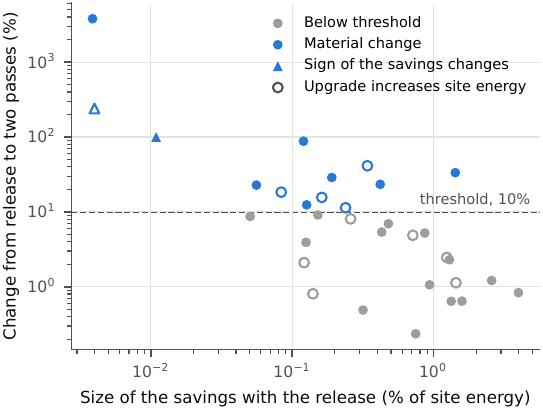}
  \caption{Absolute size of the window-upgrade savings with the release against the absolute change
  from the release to two passes, in each eligible prototype case with a nonzero change (33 of 34).}
  \label{fig:window}
\end{figure}

The savings of the window upgrade are the one outcome that changes in enough cases to matter, in 13
of 34 eligible cases, 38\% (Figure~\ref{fig:window}). In seven, the savings keep their sign and
change in size by 12 to 88\%, for example in the secondary school in Denver from 117.0 to 156.1~GJ.
In the large office in International Falls they rise from 1.1 to 44.3~GJ. In three cases in which
the upgrade increases site energy, the increase shrinks, in the large office in Denver from 102.9 to
60.4~GJ. In the primary school in Miami the savings change sign, from $-0.15$ to 0.21~GJ, and in the
secondary school in Miami they fall from 1.1~GJ to zero. In 12 of the 13 cases, the change exceeds
the spread of the savings among the three corrected configurations. An ordinary change of timestep
changes the savings about as often. With the release at four or ten timesteps per hour in place of
six, they change materially in 12 of 34 and 12 of 33 cases, and in 17 of 34 at either timestep, by a
median of 7 to 8\%, against 6\% for the correction. Ten of the 13 cases that the correction changes
materially also change materially with the timestep.

No outcome shifts in a consistent direction across the cases at the 5\% level after a Bonferroni
correction for the seven outcomes. Before that correction, cooling energy, heating capacity and
supply fan flow do, with $p = 0.009$ to $0.048$ and medians of 0.12\% or less. The verdicts depend
on where the thresholds lie. At half the thresholds, energy and autosized capacity would change
materially in 18 and 16 of 45 cases, more than a quarter, and peak demand and unmet hours in 6 each.
At twice the thresholds, only the window savings would still reach a quarter, in 9 of 34 cases. The
largest change in the window savings is 0.48\% of the building's annual site energy, in the
secondary school in Denver, and only four exceed 0.1\%.

\begin{figure}[tbp]
  \centering
  \includegraphics[width=\textwidth]{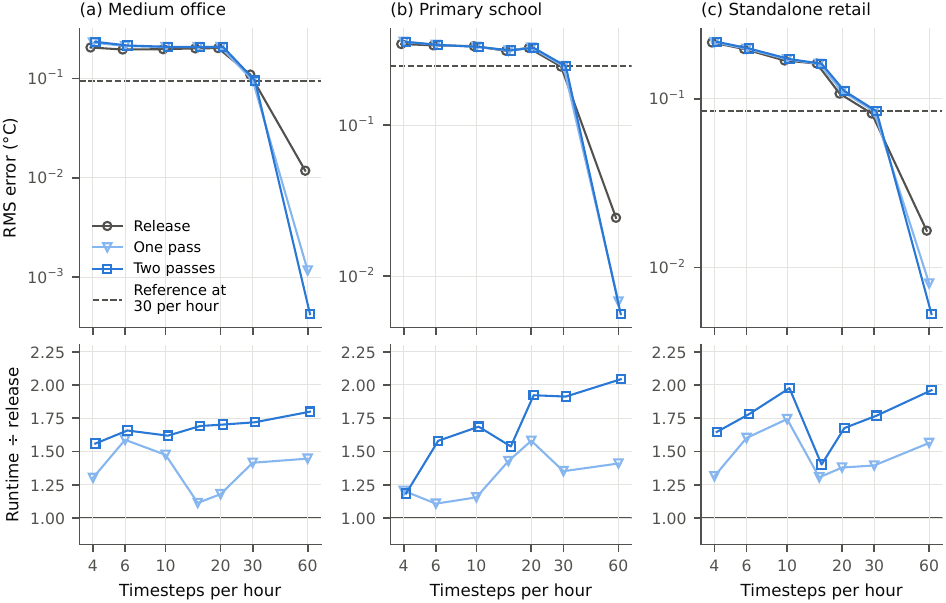}
  \caption{Error of the hourly zone air temperature against the converged configuration with 60
  timesteps per hour (top) and runtime relative to the release (bottom), at each timestep, in three
  prototype buildings in Denver. Runtime is the median of three runs, and the dashed line marks the
  converged configuration at 30 per hour.}
  \label{fig:accrt}
\end{figure}

At the six timesteps per hour of the prototype files, two passes take 1.58 to 1.78 times the runtime
of the release in the three buildings of the runtime test. They do not buy accuracy that the release
could reach only with a shorter timestep (Figure~\ref{fig:accrt}). From 4 to 20 timesteps per hour,
the error of the hourly air temperature against the converged reference stays nearly flat for all
configurations, at 0.20 to \SI{0.23}{\celsius} in the office and 0.31 to \SI{0.36}{\celsius} in the
school, and falls from 0.22 to \SI{0.11}{\celsius} in the retail building. The three configurations
differ from each other by much less than from the reference, which itself moves by 0.08 to
\SI{0.25}{\celsius} between 30 and 60 timesteps per hour. In no building does the release need twice
the runtime of two passes to reach their error.

Most of the difference from the reference comes from sizing. At six timesteps per hour, the annual
heating energy of all three configurations differs from the reference by about 9\% in the office,
8\% in the school and 19\% in the retail building, and the annual cooling energy by 2 to 6\%, with
each run sizing its equipment at its own timestep. With the sizes fixed at those of the release at
six timesteps per hour, the heating difference falls to 1.6 to 1.7\% in the school and 0.9\% in the
retail building, and the error falls steadily as the timestep shortens. The release and the
converged configuration then differ in annual heating by at most 0.24 percentage points, and the
release and two passes in hourly air temperature error by at most \SI{0.03}{\celsius}. In the
office, a heating difference of 8.4 to 9.2\% remains from 4 to 20 timesteps per hour and falls to
1.9\% at 30, in all three configurations alike, so it comes from neither the sizing nor the lag.

\section{Discussion}\label{sec:discussion}
\subsection{Why a third-order formula gives a first-order solution}\label{sec:d-order}

The measurements support H1 and H2. The release is first order or lower with all three schemes, and
solved to agreement the default formula reaches about second order. The third order that the
Engineering Reference gives for the default~\cite{eplus2026engref} thus belongs to the formula
alone. The lag limits the whole solution to first order, as the theory of coupled problems
predicts~\cite{trcka2010cosim,arnold2013error}, here inside a single program. In the room with an
exact solution, the extra heat capacity that follows from Eq.~\eqref{eq:lagerror} is 1.25~MJ/K at
six timesteps per hour, more than the 1.09~MJ/K of the air itself. The lagged room therefore
responds as if its time constant were 4.5~h in place of 2.1~h, which damps and delays the response
of the air.

As Hensen found for coupled heat and airflow models~\cite{hensen1999coupled}, a shorter timestep
buys accuracy far more slowly than agreement within the timestep does. In the room with an exact
solution, the release with 60 timesteps per hour is still five times less accurate than the
converged configuration with four.

The correction helps the three schemes unequally. The default formula gains most, because once its
inputs are current it can deliver more than first order, and where constructions have thermal mass,
conduction, which is second order, then limits it to about second order. The analytical solution
gains least, and in the room with an exact solution not at all. It integrates the air equation
exactly only while the other terms of the air balance stay constant over the step. With the balances
in agreement, these terms take their values at the end of the step, so the step behaves like a
first-order implicit method, and its description as free of truncation error~\cite{eplus2026engref}
applies to the air equation alone. The argument that the daily cycle cancels truncation
errors~\cite{taylor1991impact,eplus2026engref} did not hold either.

\subsection{Annual results and savings in the prototype buildings}\label{sec:d-proto}

The results support H3 for one of its six outcomes, the savings of the window upgrade. Two findings
explain why errors in the zone air temperature of the simple rooms became changes of at most 0.45\%
in the annual site energy of the prototype buildings. The first is thermal mass. A surface with mass
follows the air less closely, so it adds less of the extra heat flow of Eq.~\eqref{eq:lagerror}. In
the single room, replacing the concrete floor with a massless layer made the error of the default
formula at four timesteps per hour about fifteen times larger.

The second is that other timestep effects are larger than the lag in these buildings. At six
timesteps per hour, the annual heating energy of all three configurations differed from the
reference by 0.9 to 9\% with the equipment sizes fixed and by 8 to 19\% with each run sizing its
own, against a change from the correction of 0.24\% at the median and 5.1\% at most. In buildings
like these, the timestep dependence that earlier studies
reported~\cite{hong2008runtime,tabaresvelasco2013timestep} thus comes mostly from parts of the model
other than the lag.

The savings of the window upgrade are the exception. Savings are the difference between two annual
totals, each of which the correction changes by less than 0.5\%, so when the savings are small
compared with those totals, small fractions of the totals decide them. The changes exceed the spread
among the three corrected configurations, so they come from removing the lag. A change of timestep
from six to four or ten per hour changed the savings about as often, however, and mostly in the same
cases. Small savings of this kind respond to details of the calculation in general, and the lag is
one of them.

\subsection{Implications for users, code analyses and developers}\label{sec:d-impl}

For users of the release, the first-order solution gives a simple rule. Ten times the accuracy in
the zone air temperature costs at least ten times as many timesteps. The error is largest where
surfaces follow the air closely, as in zones with little thermal mass, and where the air temperature
changes quickly.

For analyses of codes and building stocks, the savings of measures are the result to check. Where
savings are small compared with the annual totals, as for the window upgrade, a second run at
another timestep or with the correction shows whether they depend on details of the calculation that
users do not normally see.

For the developers of EnergyPlus, the results suggest an option. The current calculation avoids
iteration and serves annual results well in buildings like the prototypes. For analyses in which the
response within the day matters, a setting that solves the surfaces, air and HVAC to agreement
within each timestep would let users of the third-order and Euler formulas trade runtime for
accuracy. The analytical solution would need a different update, because agreement does not reduce
its error. The setting needs no new model, only one or two further passes through the existing
calculations, at 1.35 to 1.65 times the runtime of the release in the office test and 1.6 to 1.8
times in the prototype buildings at six timesteps per hour. The Engineering Reference might also
note, next to the order of each formula, that the order of the whole solution is one or lower, and
the input guide could then tell users how fast the error falls when they shorten the timestep.

\section{Conclusions}\label{sec:conclusions}
In EnergyPlus 26.2, the surfaces of each timestep take the zone air temperature of the previous one,
and this one-step lag makes the whole solution first order or lower with all three schemes, the
third-order default included. Solving the surfaces, air and HVAC to agreement within each timestep
restores about second order for the default formula, and in the room with an exact solution it
brings the error at six timesteps per hour from 1.6 to \SI{0.025}{\celsius}. Two extra passes, at
1.65 times the runtime, reduce that error eightfold. In the 45 prototype cases, the correction
changed annual site energy by at most 0.45\%, and of the six outcomes only the savings of a window
upgrade changed materially in more than a quarter of the cases, in 13 of 34, about as often as under
a change of timestep from six to four or ten per hour. Other timestep effects, the sizing of
equipment at each timestep foremost, were larger than the lag in these buildings.

Future research can extend the tests to other room air models, to the finite-difference conduction
method and to zones with radiant systems, in which the program already recalculates the surfaces
once. It can test whether extrapolating the air temperature that the surfaces read, as co-simulation
does, gives the gain of the passes at less runtime, and trace the causes of the larger timestep
effects in the prototype buildings. Measurements in buildings whose air temperature changes quickly,
for example under pre-cooling or setpoint steps, would show how much the correction matters in
operation.

  \availability{\txtavailability}
  \funding{\txtfunding}
  \credit{\txtcredit}
  \competing{\txtcompeting}
  \aiuse{\txtaiuse}
  \appendix
  \section{Source locations of the order of calculation}\label{app:trace}
  Table~\ref{tab:trace} gives the location in the source code of each step in the order of
  calculation within a zone timestep.
  \begin{table}[htbp]
  \centering
  \caption{Locations in the source code of EnergyPlus 26.2.0 of each step in the order of
  calculation within a zone timestep.}
  \label{tab:trace}
  \begin{tabularx}{\textwidth}{@{}>{\raggedright\arraybackslash}p{0.28\textwidth}Y>{\raggedright\arraybackslash}p{0.16\textwidth}@{}}
    \toprule
    Step & Source file (routine) & Lines \\
    \midrule
    Loop over zone timesteps & \texttt{SimulationManager.cc} & 569--612 \\
    Order: outside surfaces, inside surfaces, zone air, final surface update &
      \texttt{HeatBalanceSurfaceManager.cc} \newline (\texttt{ManageSurfaceHeatBalance}) & 184, 188, 195, 200 \\
    Zone air balance calls the predictor, HVAC and corrector &
      \texttt{HeatBalanceAirManager.cc}; \newline \texttt{HVACManager.cc} \newline (\texttt{ManageHVAC}) &
      161, 4637; 265--305 \\
    Zone air temperature updated & \texttt{ZoneTempPredictorCorrector.cc} & 4175 \\
    Inside surfaces read the previous step's air temperature &
      \texttt{HeatBalanceSurfaceManager.cc} & 8186, 9045 \\
    Air temperature held fixed in the inside iteration & \texttt{HeatBalanceSurfaceManager.cc} &
      loops from 8231, 9113 \\
    Interior convection coefficients use it &
      \texttt{ConvectionCoefficients.cc} \newline (\texttt{InitIntConvCoeff}) & 291, 333 \\
    Outside faces of quick-conducting surfaces use it &
      \texttt{HeatBalanceSurfaceManager.cc} \newline (\texttt{CalcOutsideSurfTemp}) &
      9808--9821, 9936--9979 \\
    Air balance uses the current surface temperatures & \texttt{ZoneTempPredictorCorrector.cc} &
      5333--5336 \\
    Surfaces recalculated only with radiant systems and similar components &
      \texttt{HeatBalanceSurfaceManager.cc} \newline (\texttt{UpdateFinalSurfaceHeatBalance}) & 5466--5471 \\
    \bottomrule
  \end{tabularx}
  \par\smallskip\raggedright
  Lines at the release commit 4bd7a1f. Every release from 8.2.0 to 26.1.0 calculates in the same
  order, and its inside surfaces read the previous step's air temperature, for example at line 5126
  in release 8.2.0 and 7871 in 26.1.0.
\end{table}
  \HardFloatBarrier
  \section{Results for every prototype case}\label{app:cases}
  Tables~\ref{tab:case-energy}--\ref{tab:case-window} give, for every case of the prototype study,
  the change from the release to two passes in each outcome measure.
  {\setlength{\tabcolsep}{4pt}\setlength{\LTpost}{0pt}
\begin{longtable}{@{}llrrrr@{}}
  \caption{Change in annual energy from the release to two passes in each prototype case (\%).}\label{tab:case-energy} \\
  \toprule
  & Climate & Heating & Cooling & Fan & Site \\
  Building & zone & energy & energy & energy & energy \\
  \midrule
  \endfirsthead
  \multicolumn{6}{@{}l}{\emph{(continued)}} \\
  \toprule
  & Climate & Heating & Cooling & Fan & Site \\
  Building & zone & energy & energy & energy & energy \\
  \midrule
  \endhead
  \bottomrule
  \endfoot
  \bottomrule
  \endlastfoot
  High-rise apartment & 1A & (+6.96) & $-$0.10 & $-$0.20 & $-$0.07 \\
  High-rise apartment & 5B & +0.78 & $-$0.05 & +0.25 & +0.09 \\
  High-rise apartment & 7 & +0.17 & ($-$0.05) & $-$0.04 & +0.02 \\
  Mid-rise apartment & 1A & (+12.00) & $-$0.11 & $-$0.28 & $-$0.05 \\
  Mid-rise apartment & 5B & +0.51 & $-$0.37 & $-$0.14 & +0.03 \\
  Mid-rise apartment & 7 & +0.02 & ($-$0.47) & $-$0.30 & $-$0.03 \\
  Hospital & 1A & (+4.08) & +0.33 & $-$0.00 & +0.13 \\
  Hospital & 5B & +0.88 & +0.29 & +0.08 & +0.12 \\
  Hospital & 7 & +0.27 & (+0.28) & +0.01 & +0.09 \\
  Large hotel & 1A & (+5.77) & +0.04 & +0.02 & +0.01 \\
  Large hotel & 5B & +0.09 & +0.13 & +0.05 & +0.03 \\
  Large hotel & 7 & +0.17 & (+0.16) & (+0.05) & +0.06 \\
  Small hotel & 1A & (+3.03) & $-$0.01 & +0.07 & +0.01 \\
  Small hotel & 5B & +0.29 & $-$0.25 & +0.03 & +0.00 \\
  Small hotel & 7 & +0.09 & $-$0.22 & +0.03 & +0.01 \\
  Large office & 1A & (+6.71) & +0.12 & +0.88 & +0.10 \\
  Large office & 5B & +0.67 & +0.16 & +0.77 & +0.14 \\
  Large office & 7 & \textbf{$-$1.09} & (+0.23) & +0.54 & $-$0.18 \\
  Medium office & 1A & (+2.78) & +0.37 & (+0.41) & +0.15 \\
  Medium office & 5B & +0.07 & +0.30 & +0.32 & +0.06 \\
  Medium office & 7 & $-$0.73 & +0.16 & (+0.25) & $-$0.18 \\
  Outpatient health care & 1A & \textbf{$-$5.13} & +0.44 & \textbf{+1.13} & $-$0.23 \\
  Outpatient health care & 5B & \textbf{$-$2.93} & +0.55 & +0.45 & $-$0.45 \\
  Outpatient health care & 7 & \textbf{$-$1.23} & +0.93 & +0.33 & $-$0.34 \\
  Fast-food restaurant & 1A & (+1.09) & $-$0.05 & ($-$0.21) & $-$0.02 \\
  Fast-food restaurant & 5B & +0.24 & (+0.03) & (+0.05) & +0.05 \\
  Fast-food restaurant & 7 & +0.56 & ($-$0.13) & (+0.15) & +0.21 \\
  Sit-down restaurant & 1A & (+2.40) & $-$0.06 & +0.04 & $-$0.01 \\
  Sit-down restaurant & 5B & +0.05 & ($-$0.01) & +0.02 & +0.01 \\
  Sit-down restaurant & 7 & +0.07 & (+0.02) & (+0.03) & +0.03 \\
  Standalone retail & 1A & (+3.36) & +0.07 & $-$0.56 & $-$0.03 \\
  Standalone retail & 5B & +0.34 & $-$0.49 & \textbf{$-$1.11} & $-$0.08 \\
  Standalone retail & 7 & +0.02 & ($-$0.02) & \textbf{$-$1.09} & $-$0.09 \\
  Strip mall & 1A & (+3.92) & +0.08 & $-$0.13 & +0.02 \\
  Strip mall & 5B & +0.16 & $-$0.43 & $-$0.41 & $-$0.03 \\
  Strip mall & 7 & $-$0.13 & ($-$0.89) & +0.82 & $-$0.05 \\
  Primary school & 1A & (+1.10) & +0.80 & +0.85 & +0.38 \\
  Primary school & 5B & $-$0.13 & +0.03 & +0.41 & +0.00 \\
  Primary school & 7 & +0.03 & ($-$0.04) & +0.44 & +0.04 \\
  Secondary school & 1A & (+0.32) & +0.35 & +0.31 & +0.20 \\
  Secondary school & 5B & \textbf{+1.13} & +0.14 & $-$0.96 & +0.14 \\
  Secondary school & 7 & $-$0.17 & +0.19 & \textbf{$-$1.51} & $-$0.19 \\
  Warehouse & 1A & (+5.00) & +0.47 & +0.34 & +0.13 \\
  Warehouse & 5B & $-$0.37 & ($-$1.04) & ($-$0.13) & $-$0.24 \\
  Warehouse & 7 & $-$0.04 & (+0.00) & ($-$0.03) & $-$0.03 \\
\end{longtable}
\noindent Bold: material change (at least 1\%). Parentheses: end use below 5\% of the release's annual site energy, not counted. Every case has timesteps in which the converged configuration reached its limit of 50 passes (Table~\ref{tab:case-limits}). Climate zones: 1A Miami, 5B Denver, 7 International Falls.\par\bigskip
}
  {\setlength{\tabcolsep}{4pt}\setlength{\LTpost}{0pt}
\begin{longtable}{@{}llrrr@{}}
  \caption{Change in annual peak electric demand and autosized capacity from the release to two passes (\%), and timesteps at the pass limit of the converged configuration, in each prototype case.}\label{tab:case-limits} \\
  \toprule
  & Climate & Peak electric & Autosized & Timesteps at \\
  Building & zone & demand (\%) & capacity (\%) & pass limit \\
  \midrule
  \endfirsthead
  \multicolumn{5}{@{}l}{\emph{(continued)}} \\
  \toprule
  & Climate & Peak electric & Autosized & Timesteps at \\
  Building & zone & demand (\%) & capacity (\%) & pass limit \\
  \midrule
  \endhead
  \bottomrule
  \endfoot
  \bottomrule
  \endlastfoot
  High-rise apartment & 1A & +0.54 & $-$1.60 & 135 \\
  High-rise apartment & 5B & +0.00 & $-$1.87 & 997 \\
  High-rise apartment & 7 & +0.03 & $-$1.79 & 934 \\
  Mid-rise apartment & 1A & \textbf{+2.14} & $-$1.59 & 63 \\
  Mid-rise apartment & 5B & +1.54 & $-$1.66 & 222 \\
  Mid-rise apartment & 7 & $-$1.38 & $-$1.38 & 235 \\
  Hospital & 1A & $-$0.03 & +0.05 & 37,382 \\
  Hospital & 5B & $-$0.14 & +0.11 & 34,535 \\
  Hospital & 7 & +0.17 & $-$0.11 & 39,214 \\
  Large hotel & 1A & $-$0.05 & +0.11 & 1,104 \\
  Large hotel & 5B & $-$0.03 & +0.14 & 2,868 \\
  Large hotel & 7 & $-$0.33 & +0.18 & 7,464 \\
  Small hotel & 1A & $-$0.14 & +0.02 & 121 \\
  Small hotel & 5B & +0.06 & +0.02 & 764 \\
  Small hotel & 7 & +0.01 & +0.02 & 952 \\
  Large office & 1A & +0.34 & $-$1.23 & 4,674 \\
  Large office & 5B & $-$0.26 & $-$1.12 & 3,806 \\
  Large office & 7 & +0.56 & $-$1.15 & 9,798 \\
  Medium office & 1A & +0.09 & $-$0.26 & 1,613 \\
  Medium office & 5B & $-$1.86 & $-$0.27 & 154 \\
  Medium office & 7 & $-$1.08 & $-$0.22 & 1,956 \\
  Outpatient health care & 1A & +0.41 & $-$0.84 & 20,706 \\
  Outpatient health care & 5B & +0.70 & $-$1.24 & 27,963 \\
  Outpatient health care & 7 & +0.44 & $-$1.21 & 31,812 \\
  Fast-food restaurant & 1A & +0.02 & $-$0.04 & 3,426 \\
  Fast-food restaurant & 5B & +0.00 & $-$0.01 & 81 \\
  Fast-food restaurant & 7 & $-$0.04 & $-$0.01 & 42 \\
  Sit-down restaurant & 1A & +0.00 & +0.00 & 777 \\
  Sit-down restaurant & 5B & $-$0.00 & $-$0.01 & 37 \\
  Sit-down restaurant & 7 & +0.02 & $-$0.02 & 30 \\
  Standalone retail & 1A & +0.05 & +1.89 & 8,777 \\
  Standalone retail & 5B & $-$0.08 & $-$1.12 & 2,382 \\
  Standalone retail & 7 & $-$0.10 & $-$1.01 & 1,792 \\
  Strip mall & 1A & $-$0.51 & $-$0.13 & 13 \\
  Strip mall & 5B & +1.49 & $-$0.58 & 854 \\
  Strip mall & 7 & +0.95 & +1.03 & 778 \\
  Primary school & 1A & +0.04 & +1.92 & 5,639 \\
  Primary school & 5B & +0.14 & +0.08 & 4,334 \\
  Primary school & 7 & +0.25 & +0.08 & 8,850 \\
  Secondary school & 1A & +0.04 & $-$0.38 & 6,664 \\
  Secondary school & 5B & $-$0.38 & $-$0.46 & 5,334 \\
  Secondary school & 7 & $-$0.02 & $-$0.66 & 15,229 \\
  Warehouse & 1A & +0.21 & +0.00 & 34 \\
  Warehouse & 5B & +0.02 & +0.01 & 129 \\
  Warehouse & 7 & $-$0.00 & +0.01 & 29 \\
\end{longtable}
\noindent Bold: material change (at least 2\%). Autosized capacity: the largest change among the sums of cooling coil, heating coil and supply fan capacities. Timesteps at pass limit: timesteps of the year, out of 52,560, in which the converged configuration reached its limit of 50 passes. Climate zones: 1A Miami, 5B Denver, 7 International Falls.\par\bigskip
}
  {\setlength{\tabcolsep}{4pt}\setlength{\LTpost}{0pt}
\begin{longtable}{@{}llrrrr@{}}
  \caption{Occupied unmet heating and cooling hours in each prototype case, release and two passes.}\label{tab:case-unmet} \\
  \toprule
  & Climate & \multicolumn{2}{c}{Heating (h)} & \multicolumn{2}{c}{Cooling (h)} \\
  \cmidrule(lr){3-4}\cmidrule(l){5-6}
  Building & zone & Release & Two passes & Release & Two passes \\
  \midrule
  \endfirsthead
  \multicolumn{6}{@{}l}{\emph{(continued)}} \\
  \toprule
  & Climate & \multicolumn{2}{c}{Heating (h)} & \multicolumn{2}{c}{Cooling (h)} \\
  \cmidrule(lr){3-4}\cmidrule(l){5-6}
  Building & zone & Release & Two passes & Release & Two passes \\
  \midrule
  \endhead
  \bottomrule
  \endfoot
  \bottomrule
  \endlastfoot
  High-rise apartment & 1A & 0.0 & 0.0 & 5.0 & 5.0 \\
  High-rise apartment & 5B & 0.0 & 0.0 & 6.3 & 7.3 \\
  High-rise apartment & 7 & 0.0 & 0.0 & 5.2 & 6.5 \\
  Mid-rise apartment & 1A & 0.0 & 0.0 & 0.0 & 0.0 \\
  Mid-rise apartment & 5B & 0.0 & 0.0 & 0.0 & 0.0 \\
  Mid-rise apartment & 7 & 0.0 & 0.0 & 0.2 & 0.2 \\
  Hospital & 1A & 0.0 & 0.0 & 885.0 & 895.7 \\
  Hospital & 5B & 0.0 & 0.0 & 62.5 & 68.7 \\
  Hospital & 7 & 19.3 & 18.7 & 13.2 & 11.5 \\
  Large hotel & 1A & 0.0 & 0.0 & \textbf{78.5} & \textbf{69.7} \\
  Large hotel & 5B & 12.8 & 12.0 & 488.8 & 483.3 \\
  Large hotel & 7 & 315.3 & 316.2 & 45.5 & 42.0 \\
  Small hotel & 1A & 0.0 & 0.0 & 343.7 & 332.3 \\
  Small hotel & 5B & 4.0 & 3.8 & 240.8 & 245.3 \\
  Small hotel & 7 & 13.3 & 11.2 & 464.8 & 470.7 \\
  Large office & 1A & 0.0 & 0.0 & 640.0 & 677.0 \\
  Large office & 5B & 64.7 & 66.7 & 645.0 & 677.8 \\
  Large office & 7 & 315.2 & 307.5 & 319.7 & 322.2 \\
  Medium office & 1A & 7.2 & 7.5 & 511.0 & 525.2 \\
  Medium office & 5B & 206.0 & 206.3 & 184.8 & 187.3 \\
  Medium office & 7 & 627.2 & 609.8 & 87.5 & 89.2 \\
  Outpatient health care & 1A & 335.3 & 334.5 & 2535.8 & 2538.7 \\
  Outpatient health care & 5B & 2436.2 & 2440.8 & 842.2 & 841.7 \\
  Outpatient health care & 7 & 3157.2 & 3159.2 & 482.7 & 483.5 \\
  Fast-food restaurant & 1A & 0.0 & 0.0 & 930.8 & 935.0 \\
  Fast-food restaurant & 5B & 0.0 & 0.2 & 0.2 & 0.3 \\
  Fast-food restaurant & 7 & 0.2 & 0.2 & 7.5 & 7.7 \\
  Sit-down restaurant & 1A & 0.0 & 0.0 & 890.5 & 895.7 \\
  Sit-down restaurant & 5B & 0.0 & 0.2 & 0.0 & 0.0 \\
  Sit-down restaurant & 7 & 0.2 & 0.2 & 4.2 & 4.2 \\
  Standalone retail & 1A & 5.7 & 5.7 & 0.0 & 0.0 \\
  Standalone retail & 5B & 332.7 & 334.0 & 0.0 & 0.0 \\
  Standalone retail & 7 & 334.7 & 333.7 & 0.0 & 0.0 \\
  Strip mall & 1A & 0.0 & 0.0 & 0.0 & 0.0 \\
  Strip mall & 5B & 1.2 & 1.2 & 0.0 & 0.0 \\
  Strip mall & 7 & 9.0 & 9.0 & 0.0 & 0.0 \\
  Primary school & 1A & 25.8 & 22.5 & 1050.7 & 1058.8 \\
  Primary school & 5B & 721.2 & 705.7 & 111.2 & 112.2 \\
  Primary school & 7 & 1449.3 & 1433.8 & 37.3 & 37.5 \\
  Secondary school & 1A & 64.8 & 64.0 & 1760.8 & 1755.0 \\
  Secondary school & 5B & 1291.8 & 1278.3 & 264.0 & 267.0 \\
  Secondary school & 7 & 1890.5 & 1880.5 & 83.8 & 83.8 \\
  Warehouse & 1A & 0.2 & 0.0 & 58.3 & 58.7 \\
  Warehouse & 5B & 69.5 & 66.8 & 2.0 & 1.8 \\
  Warehouse & 7 & 98.8 & 98.5 & 0.8 & 1.0 \\
\end{longtable}
\noindent Bold: material change (at least 10\% and 5 h for heating or cooling, or the total crossing 300 h). Every case has timesteps in which the converged configuration reached its limit of 50 passes (Table~\ref{tab:case-limits}). Climate zones: 1A Miami, 5B Denver, 7 International Falls.\par\bigskip
}
  {\setlength{\tabcolsep}{4pt}\setlength{\LTpost}{0pt}
\begin{longtable}{@{}llrrr@{}}
  \caption{Annual site energy savings of the window upgrade in each eligible prototype case, release and two passes.}\label{tab:case-window} \\
  \toprule
  & Climate & \multicolumn{2}{c}{Savings (GJ)} & Change \\
  \cmidrule(lr){3-4}
  Building & zone & Release & Two passes & (\%) \\
  \midrule
  \endfirsthead
  \multicolumn{5}{@{}l}{\emph{(continued)}} \\
  \toprule
  & Climate & \multicolumn{2}{c}{Savings (GJ)} & Change \\
  \cmidrule(lr){3-4}
  Building & zone & Release & Two passes & (\%) \\
  \midrule
  \endhead
  \bottomrule
  \endfoot
  \bottomrule
  \endlastfoot
  High-rise apartment & 5B & 162.9 & 164.2 & +0.8 \\
  High-rise apartment & 7 & 49.8 & 50.3 & +1.1 \\
  Mid-rise apartment & 5B & 36.3 & 36.7 & +1.2 \\
  Mid-rise apartment & 7 & 9.6 & 9.6 & +0.0 \\
  Hospital & 5B & 121.3 & 129.8 & +6.9 \\
  Hospital & 7 & $-$76.3 & $-$67.7 & \textbf{$-$11.4} \\
  Large hotel & 5B & 173.8 & 174.9 & +0.6 \\
  Large hotel & 7 & 15.5 & 17.4 & \textbf{+12.4} \\
  Small hotel & 5B & 21.2 & 21.2 & +0.2 \\
  Small hotel & 7 & 10.2 & 10.2 & $-$0.5 \\
  Large office & 5B & $-$102.9 & $-$60.4 & \textbf{$-$41.3} \\
  Large office & 7 & 1.1 & 44.3 & \textbf{+3788.6} \\
  Medium office & 5B & 2.1 & 3.9 & \textbf{+87.9} \\
  Medium office & 7 & $-$1.6 & $-$1.3 & \textbf{$-$18.4} \\
  Outpatient health care & 5B & 54.6 & 55.0 & +0.6 \\
  Outpatient health care & 7 & 21.2 & 22.3 & +5.4 \\
  Fast-food restaurant & 5B & 0.7 & 0.8 & +8.7 \\
  Fast-food restaurant & 7 & $-$2.5 & $-$2.5 & $-$0.8 \\
  Sit-down restaurant & 5B & 1.1 & 1.4 & \textbf{+22.8} \\
  Sit-down restaurant & 7 & $-$3.3 & $-$3.3 & $-$2.1 \\
  Standalone retail & 1A & $-$3.1 & $-$2.9 & $-$8.1 \\
  Standalone retail & 5B & 1.5 & 1.6 & +3.9 \\
  Standalone retail & 7 & $-$10.7 & $-$10.2 & $-$4.9 \\
  Strip mall & 5B & $-$15.0 & $-$14.7 & $-$2.5 \\
  Strip mall & 7 & $-$26.5 & $-$26.2 & $-$1.1 \\
  Primary school & 1A & $-$0.1 & 0.2 & \textbf{$-$240.0} \\
  Primary school & 5B & 43.5 & 44.5 & +2.3 \\
  Primary school & 7 & 18.4 & 22.7 & \textbf{+23.4} \\
  Secondary school & 1A & 1.1 & 0.0 & \textbf{$-$100.0} \\
  Secondary school & 5B & 117.0 & 156.1 & \textbf{+33.4} \\
  Secondary school & 7 & 87.8 & 83.2 & $-$5.2 \\
  Warehouse & 1A & 0.7 & 0.7 & +9.1 \\
  Warehouse & 5B & $-$1.4 & $-$1.6 & \textbf{+15.6} \\
  Warehouse & 7 & 3.4 & 4.4 & \textbf{+28.8} \\
\end{longtable}
\noindent Bold: material change (at least 10\%, or a change of sign). Negative savings: the upgrade increases annual site energy. Change: (two passes $-$ release) $\div$ release. Where the release's savings are negative, a positive change means the increase in energy use grew. Every case has timesteps in which the converged configuration reached its limit of 50 passes (Table~\ref{tab:case-limits}). Climate zones: 1A Miami, 5B Denver, 7 International Falls.\par\bigskip
}

\end{document}